\documentclass{PoS}
\newcommand{\rmn}{\mathrm}

\title{Neutrinos from Clusters of Galaxies and Radio Constraints}

\ShortTitle{Neutrinos from Clusters of Galaxies and Radio Constraints}

\author{\speaker{Fabio Zandanel}\\
        GRAPPA Institute, University of Amsterdam, Science Park 904, 1098 XH Amsterdam, Netherland\\
        E-mail: \email{f.zandanel@uva.nl}}
\author{Irene Tamborra\\
        GRAPPA Institute, University of Amsterdam, Science Park 904, 1098 XH Amsterdam, Netherland\\
        E-mail: \email{i.tamborra@uva.nl}}
\author{Stefano Gabici\\
        APC, Univ. Paris Diderot, CNRS/IN2P3, CEA/Irfu, Obs. de Paris, Sorbonne Paris Cit\'{e}, France\\
        E-mail: \email{gabici@apc.in2p3.fr}}
\author{Shin'ichiro Ando\\
        GRAPPA Institute, University of Amsterdam, Science Park 904, 1098 XH Amsterdam, Netherland\\
        E-mail: \email{s.ando@uva.nl}}

\abstract{Cosmic-ray (CR) protons can accumulate for cosmological times in clusters of galaxies. Their hadronic interactions with protons of the intra-cluster medium (ICM) generate secondary electrons, gamma-rays and high-energy neutrinos. In light of the high-energy neutrino events recently discovered by the IceCube observatory, we estimate the contribution from galaxy clusters to the diffuse gamma-ray and neutrino backgrounds. For the first time, we consistently take into account the synchrotron emission generated by secondary electrons and require the clusters radio counts to be respected. For a choice of parameters respecting current constraints from radio to gamma-rays, and assuming a proton spectral index of $-2$, we find that hadronic interactions in clusters contribute by less than $10$\% to the IceCube flux, and much less to the total extragalactic gamma-ray background observed by \emph{Fermi}. They account for less than $1$\% for spectral indexes $<-2$. The high-energy neutrino flux observed by IceCube can be reproduced without violating radio constraints only if a very hard (and speculative) spectral index $>-2$ is adopted. However, this scenario is in tension with the high-energy IceCube data, which seem to suggest a spectral energy distribution of the neutrino flux that decreases with the particle energy. We stress that our results are valid for all kind of sources injecting CR protons into the ICM, and that, while IceCube can test the most optimistic scenarios for spectral indexes $\geq-2.2$ by stacking few nearby massive objects, clusters of galaxies cannot give any relevant contribution to the extragalactic gamma-ray and neutrino backgrounds in any realistic scenario.}

\FullConference{The 34th International Cosmic Ray Conference,\\
		30 July- 6 August, 2015\\
		The Hague, The Netherlands}

\begin{document}

%%%%%%%%%%%%%%%%%%%%%%%%%%%%%%%%%%%%%%%%%%%%%%%%%%%%%%%%
%%%%%%%%%%%%%%%%%%%%%%%%%%%%%%%%%%%%%%%%%%%%%%%%%%%%%%%%
\section{Introduction}
The IceCube neutrino observatory at the South Pole has recently reported evidence of
extraterrestrial neutrinos~\cite{1}. These neutrinos are compatible with a flux isotropically distributed in the 
sky, with astrophysical origin and with a possible cutoff at a few PeV. The origin of these events is unknown, however, 
the isotropic distribution in the sky of the observed events suggests that they might come from various extragalactic 
$\sim100$~PeV cosmic-ray (CR) accelerators (see, e.g., \cite{2}).

As discussed in Ref.~\cite{3}, a multi-messenger connection between the measured neutrino 
fluxes and their photon counterparts could be crucial for unveiling the origin of the high-energy neutrinos. We 
here assume that the IceCube high-energy neutrinos have an extragalactic diffuse origin and are produced in proton-proton 
collisions. Therefore, they should be accompanied by gamma rays at a flux comparable to that of neutrinos. 
The extragalactic gamma-ray background (EGB) is the measured radiation that remains after subtracting all known sources from the 
observed gamma-ray sky, and it is due to the sum of contributions from different unresolved sources (see, e.g., \cite{4}). If the 
IceCube neutrinos are indeed produced in proton-proton interactions, the same sources should also contribute to the EGB.

Clusters of galaxies are the latest and largest structures to form in the Universe, where energies of 
the same order of magnitude as the gravitational binding energy, $10^{61}$--$10^{63}$~erg, are dissipated through 
structure-formation shocks and turbulence \cite{5}. Even if only a small part of this energy goes 
into particle acceleration, clusters should host significant non-thermal emission (see, e.g., \cite{6}). 
The contribution of clusters of galaxies to the EGB has been discussed by several authors \cite{7,coma_me}, 
and it has been argued that CR hadronic interactions, with the protons of the intra-cluster medium (ICM), in galaxy clusters could be responsible for a relevant fraction 
of the IceCube neutrinos \cite{3,murase2008,8}. However, such hadronic interactions could have a dramatic impact on the radio frequencies
since secondary electrons are also produced in proton-proton interactions and radiate synchrotron 
emission when interacting with the magnetic fields in clusters of galaxies (see, e.g., \cite{9}). 
Therefore, the radio emission from secondary electrons needs to respect radio counts of galaxy clusters
\cite{10}. 

In this proceeding, we summarise the main findings published in our recent article \cite{11} where we estimated 
the possible contribution to the extragalactic gamma-ray and neutrino backgrounds from galaxy clusters 
assuming that gamma rays and neutrinos mainly originate in proton-proton interactions, 
while for the first time taking the constraints from the radio regime into account. 

%%%%%%%%%%%%%%%%%%%%%%%%%%%%%%%%%%%%%%%%%%%%%%%%%%%%%%%%
%%%%%%%%%%%%%%%%%%%%%%%%%%%%%%%%%%%%%%%%%%%%%%%%%%%%%%%%
\section{Methods}
The total gamma-ray intensity from all galaxy clusters in the Universe at a given energy (d$N$\,/\,d$A$\,d$t$\,d$E$) is given by
\begin{eqnarray}
I_{\gamma} = \int_{z_1}^{z_2} \int_{M_{500,\,\rmn{lim}}} & &
 \frac{L_{\gamma}(M_{500},z)\,(1+z)^2}{4\pi D_{\rm L}(z)^2} \\ \nonumber
& \times & \frac{\rmn{d}^{2}n (M_{500},z)}{\rmn{d}V_{\rmn{c}} \, \rmn{d}M_{500}} \frac{\rmn{d}V_{\rmn{c}}}{\rmn{d}z} \rmn{d}z \, \rmn{d}M_{500}  \, ,
\label{eq:TOTgamma}
\end{eqnarray}
where the cluster mass $M_{\Delta}$ is defined with respect to a density that is $\Delta=500$ times 
the \emph{\emph{critical}} density of the Universe at redshift $z$. Here,
$L_{\gamma}$ is the gamma-ray luminosity of clusters, $V_{\rmn{c}}$ is the comoving volume, $D_{\rm L}(z)$  the luminosity distance, and 
$\rmn{d}^{2}n (M_{500},z)/\rmn{d}V_{\rmn{c}} \, \rmn{d}M_{500}$ is the cluster mass function. 
The lower limit of the mass integration has been chosen to be $M_{500,\,{\rm lim}} = 10^{13.8}$~$h^{-1}$~M$_{\odot}$, 
to account for large galaxy groups. The redshift integration goes from $z_1 = 0.01$, where the closest galaxy clusters 
are located, up to $z_2 = 2$. At the same time, the total number of detectable galaxy clusters in the radio regime at 
$f = 1.4$~GHz, above the flux limit $F_{\rmn{min}}$, is obtained as
\begin{equation}
N_{1.4} (>F_{\rmn{min}}) = \int_{z_1}^{z_2} \int_{F_{\rmn{min}}}^{\infty} \frac{\rmn{d}^{2}n (F_{1.4},z)}{\rmn{d}V_{\rmn{c}} \, \rmn{d}F_{1.4}} \frac{\rmn{d}V_{\rmn{c}}}{\rmn{d}z} \rmn{d}z \, \rmn{d}F_{1.4}  \, ,
\label{eq:Nradio}
\end{equation}
where $F_{1.4} = L_{1.4} (1+z) / 4 \pi D_{\rm L}(z)^2$, with $L_{1.4}$ the radio luminosity at 1.4~GHz. 
We compare this number with the radio counts from the 
National Radio Astronomy Observatory Very Large Array sky survey (NVSS) from Ref.~\cite{12}. 
The flux $F_{\rmn{min}}$ is defined as in equation~(9) of \cite{13} by adopting a noise-level multiplier
$\xi_1 = 1$, which is appropriate, while slightly optimistic, for the low redshifts of the NVSS survey 
($0.44\leq z \leq 0.2$), and a typical radio half-light radius of $R_{500}/4$ \cite{14}. 

The function $\rmn{d}^{2}n (F_{1.4},z)/\rmn{d}V_{\rmn{c}} \, \rmn{d}F_{1.4}$ is obtained numerically from 
$\rmn{d}^{2}n (M_{500},z)/\rmn{d}V_{\rmn{c}} \, \rmn{d}M_{500}$ by calculating $L_{1.4}(M_{500})$ from
$L_{\gamma}(M_{500})$ as explained in the following. We introduce a 
phenomenological gamma-ray luminosity-mass relation:
\begin{equation}
\rmn{log}_{10} \left[\frac{L_{\gamma} (100\,\rmn{MeV})}{\mathrm{s^{-1}\,GeV^{-1}}}\right] = P_{1} + P_{2} \, \rmn{log}_{10}
\left(\frac{M_{500}}{\mathrm{M}_\odot}\right) \, ,
\label{eq:LM}
\end{equation}
where we omit the possible redshift-dependence for simplicity. The radio luminosity can be obtained from the gamma-ray 
one by using the ``standard'' formulae for the hadronic proton-proton interactions for which we adopt  
Ref.~\cite{15}, considering a power law in momentum, $p^{-\alpha_\rmn{p}}$, for the spectral distribution of CR protons in clusters. 
In order to do so, we also assume that the magnetic field $B$ in galaxy clusters is independent of the radius in the radio-emitting region.

We note that the parameters $P_1$, $P_2$, $\alpha_{\rm p}$, $B$, and the fraction of loud clusters are degenerate when 
one tries to find the maximum allowed hadronic-induced emission. The most recent estimates suggest that the percentage of
radio-emitting (loud) clusters is about 20--30\%~\cite{16}. The subdivision of the cluster population into radio-loud and radio-quiet 
clusters is also reflected in the corresponding gamma-ray and neutrino fluxes. Therefore, from now on we refer to the two populations 
as ``loud'' and ``quiet'', while we will mainly consider the overly optimistic case where all the clusters are loud (100\% loud). In the following, 
to reduce the number of free parameters, we fix $P_2 = 5/3 \simeq 1.67$; i.e., we assume that the hadronic-induced luminosity scales as the 
cluster thermal energy $E_{\rm th} \propto M^2/R_{\rm vir} \propto M^{5/3}$, where $R_{\rm vir}$ is the virial radius. The parameter $P_1$ is 
then found requiring to not overshoot the radio counts from the NVSS survey and/or current gamma-ray upper limits, where for the latter
we take the Coma cluster as reference case \cite{coma_me}. Eventually, we also calculate the total neutrino flux from all galaxy clusters in 
the Universe, similarly to what done in Eq.~(\ref{eq:TOTgamma}), as prescribed in Ref.~\cite{17}.

We investigate different spectral indexes $\alpha_{\rmn{p}} = 1.5$, 2, 2.2, and 2.4, and different magnetic field values $B \gg B_{\rmn{CMB}}$,\footnote{$B_\mathrm{CMB} \simeq 3.2 \times (1 + z)^2$~$\mu$G is the equivalent CMB magnetic field strength such that $B_\mathrm{CMB}^2/8\pi$ is the cosmic microwave background energy density.} 
$B = 1$~$\mu$G, and $0.5$~$\mu$G. The first choice of the magnetic field can be regarded as conservative considering that the volume-averaged 
magnetic field of Coma, the best-studied cluster for Faraday rotation measurements, is about 2~$\mu$G \cite{18}; while the 
last should be considered optimistic with respect to current estimates. To clarify the meaning of the terms conservative/optimistic, note that the 
higher the magnetic field, the less room there is for protons, because the radio counts have to be respected, hence the lower the gamma-ray 
and neutrino fluxes.

\vspace {0.2cm}

In addition to the above phenomenological method based on the luminosity--mass relation, we also adopt a more sophisticated approach for modelling
the CR, ICM, and magnetic field distributions in galaxy clusters following Refs.~\cite{11,23}. In this semi-analytical approach, the galaxy cluster 
population, as well as the ICM and magnetic field distributions, are modelled such that they correctly reproduce on average current observations. 
For the CR spatial and spectral distribution, we adopt the hadronic model proposed in Ref.~\cite{11}, which extends the semi-analytical 
model of Ref.~\cite{24} to account for possible CR transport phenomena.

%%%%%%%%%%%%%%%%%%%%%%%%%%%%%%%%%%%%%%%%%%%%%%%%%%%%%%%%
%%%%%%%%%%%%%%%%%%%%%%%%%%%%%%%%%%%%%%%%%%%%%%%%%%%%%%%%
\section{Results}
\begin{figure*}[hbt!]
\centering
\includegraphics[width=0.457\textwidth]{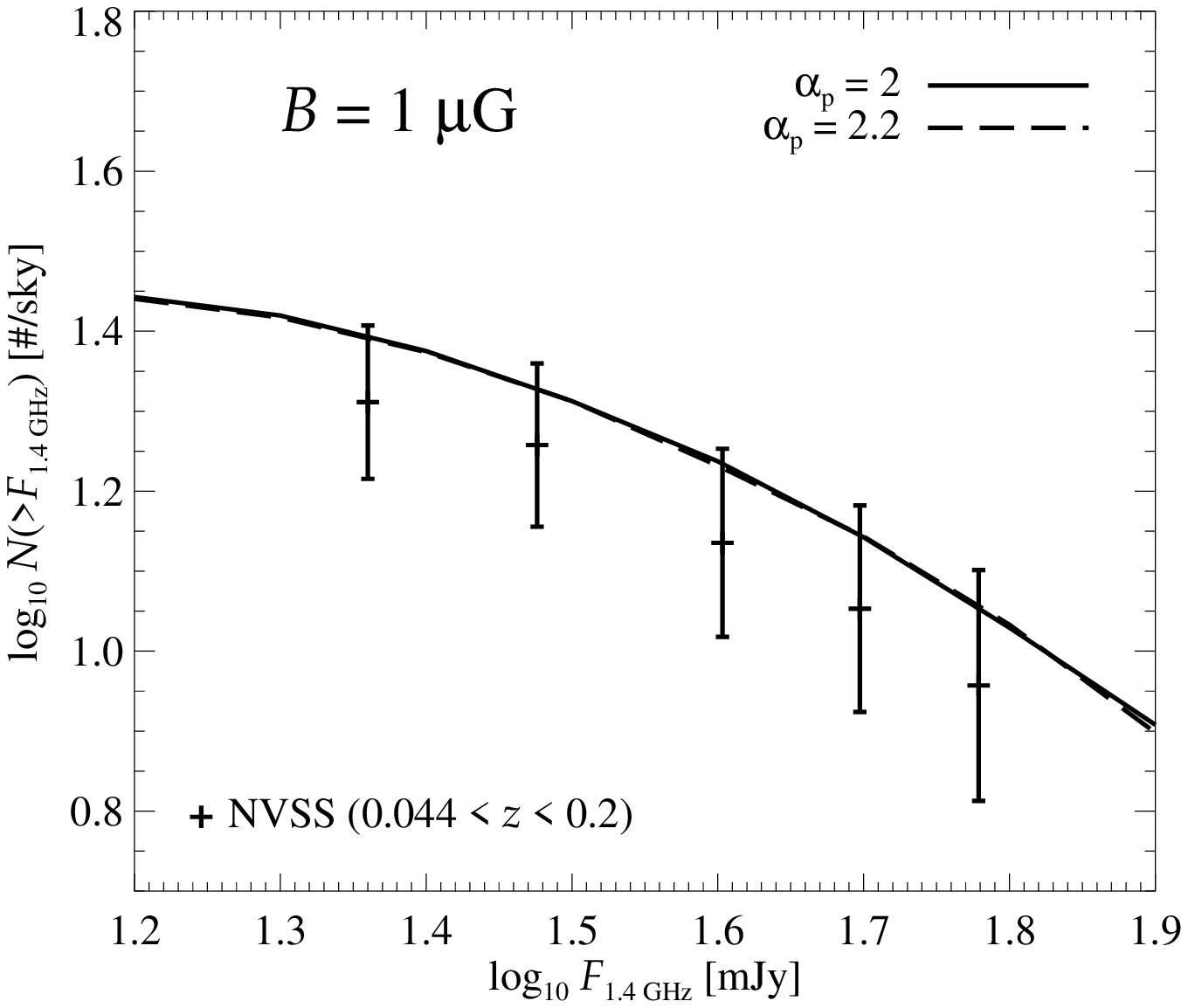}
\includegraphics[width=0.536\textwidth]{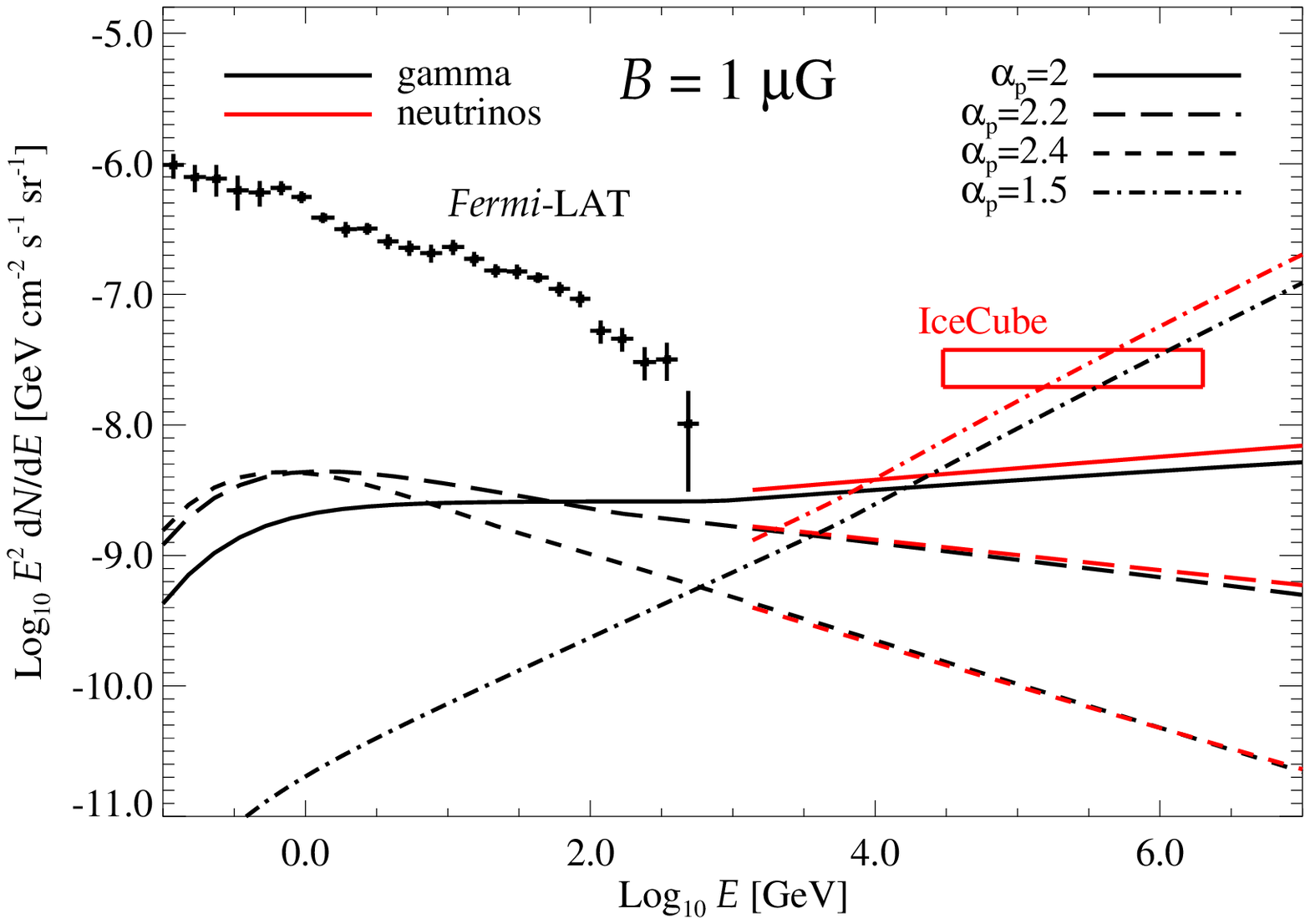}
\caption{Total gamma-ray and neutrino intensities (right) due to hadronic interactions in galaxy clusters and the corresponding radio counts due to 
synchrotron emission from secondary electrons (left) for 100\% loud clusters and $B=1$~$\mu$G obtained with the luminosity--mass relation approach. 
The neutrino intensity is meant for all flavours. All the plotted intensities respect NVSS radio counts and the gamma-ray upper limits on individual clusters. 
For $\alpha_{\rmn{p}} = 1.5$ and 2.4, the radio counts respecting the gamma-ray and neutrino limits, respectively, are below the y-scale range adopted 
for the panel on the left.}
\label{fig:resultsLM1}
\end{figure*}

While we point the reader to our article \cite{11} for a more detailed account, we summarise here our main findings. Figure~\ref{fig:resultsLM1} 
shows the results obtained with the luminosity--mass relation approach for $B=1$~$\mu$G, which we consider the most realistic choice, for both the 
comparison of our models to the radio counts (on the left) and for the computed gamma-ray (in black) and neutrino intensities (in red) as functions of the 
energy (on the right), for the chosen values of $\alpha_{\rmn{p}}$ assuming 100\% loud clusters. For comparison, we plot the {\it Fermi}-Large Area 
Telescope (LAT) data \cite{19}, and the four-years IceCube 1~$\sigma$ error band \cite{1}.

\begin{figure*}[hbt!]
\centering
\includegraphics[width=0.457\textwidth]{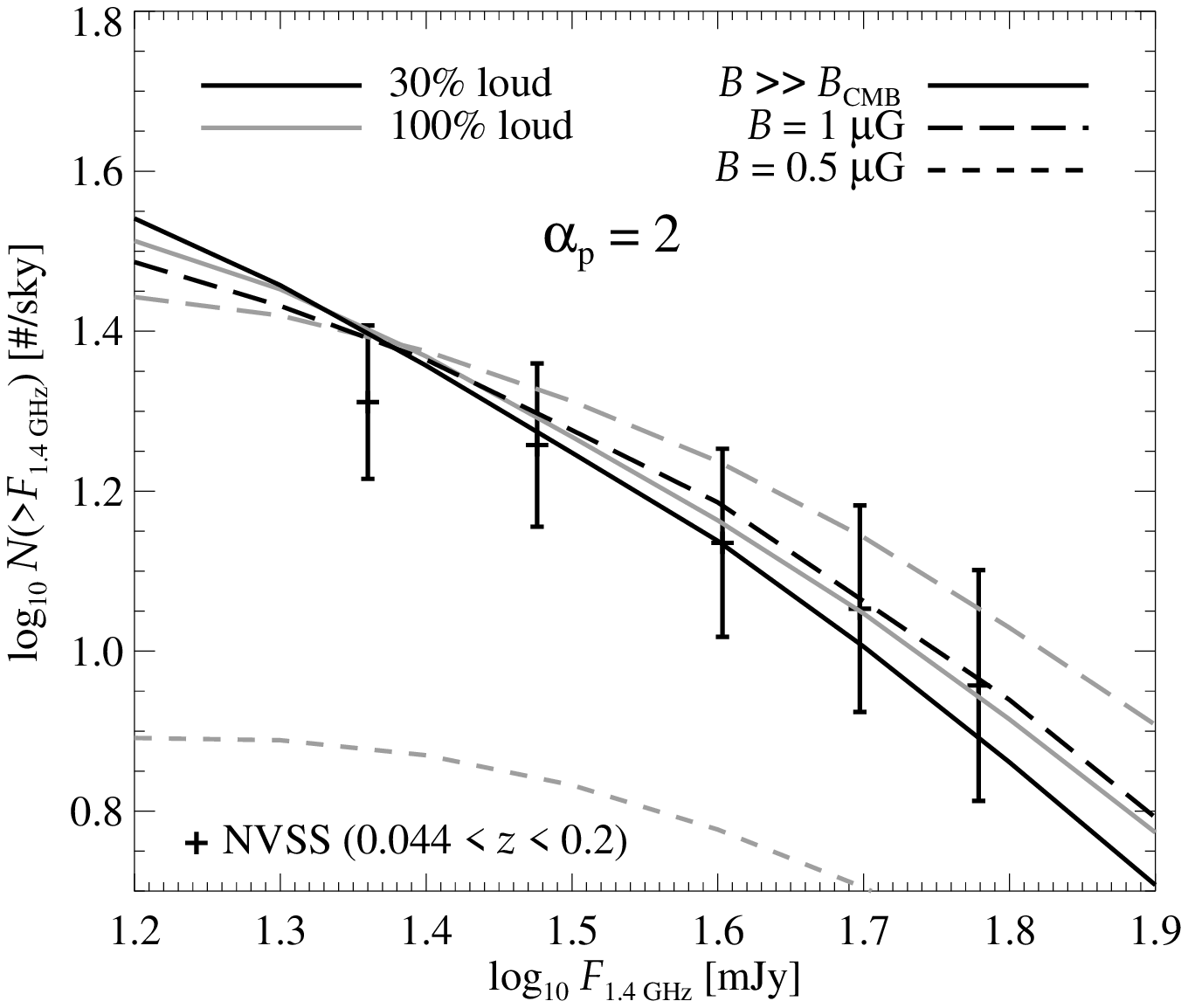}
\includegraphics[width=0.536\textwidth]{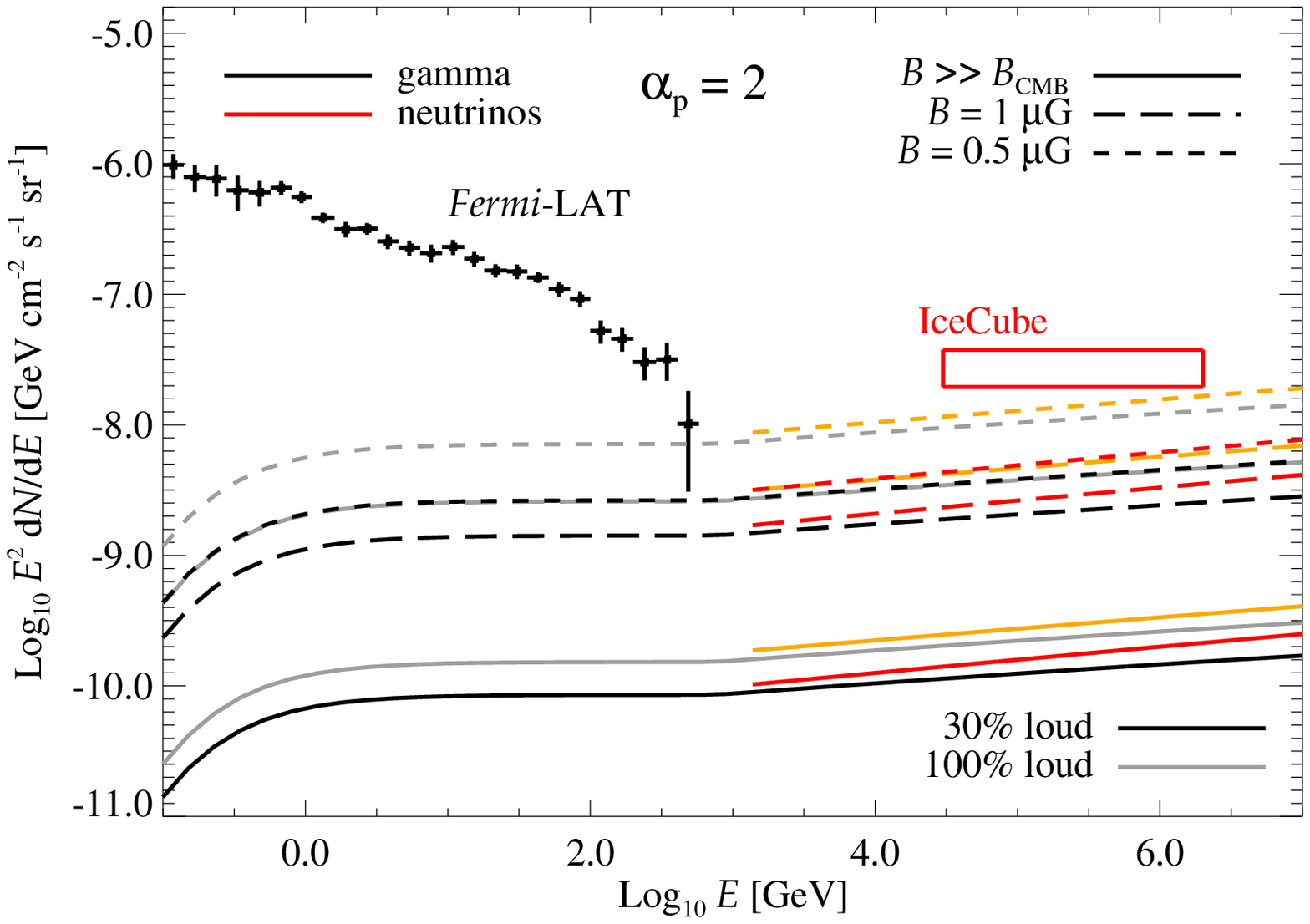}
\caption{
Same as Figure~\protect\ref{fig:resultsLM2}, together with the case of 30\% loud clusters for $\alpha_{\rmn{p}} = 2$. 
The remaining percentage of $70\%$ quiet clusters has been assumed to have $L_{\gamma} (100\,\rmn{MeV})$ one order of 
magnitude lower than for the loud clusters. The 100\% loud case is shown with lighter colours (i.e., in grey and orange).
}
\label{fig:resultsLM2}
\end{figure*}

For $\alpha_\rmn{p} > 2$, both the gamma-ray and the neutrino diffuse backgrounds are well below the {\it Fermi} 
and the IceCube data in all cases. For $\alpha_\rmn{p} = 2$, while the gamma-ray flux is always lower 
than the {\it Fermi} measurements, the neutrino diffuse background could represent a significant fraction of
the flux measured by IceCube (for the cases of $B = 1$~$\mu$G and 0.5~$\mu$G). However, the case 
of 100\% loud clusters is not realistic. Therefore, in Figure~\ref{fig:resultsLM2}, we show the same as in 
Figure~\ref{fig:resultsLM1} for $\alpha_\rmn{p} = 2$ only, together with the more realistic case of 30\% loud 
clusters. In the latter, galaxy clusters could make up at most about 10\% of the neutrino flux measured 
by Ice Cube for $B = 1$~$\mu$G. 

In the extreme case of $\alpha_\rmn{p} = 1.5$, we could explain the IceCube data by averaging over the corresponding energies
for all cases, while respecting all other constraints from radio to gamma rays. However, we note that such a hard spectral index contradicts
the most recent IceCube results suggesting a much softer spectral index \cite{1}.

\vspace{0.2cm}

Let us underline some of the assumptions made in obtaining these estimations. We do not assume any CR spectral cut-off or steepening 
possibly due to the high-energy protons that are no longer confined to the cluster \cite{20,24}. While this is not relevant when comparing with the \emph{Fermi} 
data, it might be relevant for the high-energy neutrino flux and, therefore, our results should be considered conservative in this sense. Additionally, 
we omit the absorption of high-energy gamma rays due to the interaction with the extragalactic background light because this 
becomes relevant at energies above few hundred GeV for low redshift sources such as galaxy clusters \cite{21}. Another important
assumption, worth discussing here, is the slope of the luminosity-mass relation $P_2 = 5/3$. While this choice is appropriate for CR protons injected by 
structure formation shocks, it could be different for other CR sources. We, however, estimate that a much steeper slope would 
imply lower gamma-ray and neutrino fluxes, as it would require to lower $P_1$ to not overshoot radio counts, and that a
much flatter slope could realistically boost the total extragalactic neutrino flux to be at most about 30\% of the IceCube flux 
for $\alpha_{\rmn{p}} = 2$. However, a luminosity--mass function with a slope both much steeper and much flatter than $5/3$ would
contradict current knowledge of the scaling relations of diffuse radio emission in galaxy clusters \cite{16}. 

\vspace{0.2cm}

We can conclude this section saying that amongst all the considered cases that respect both radio counts and current gamma-ray upper limits, hadronic interactions in galaxy clusters can realistically contribute at most up to $10$\% of the total extragalactic neutrino background, while contributing less than a few percentage 
points to the total extragalactic gamma-ray background. Moreover, we do know that not all the observed radio emission in clusters has a 
hadronic origin \cite{14,22}, therefore, the simplified requirement of not overshooting the NVSS radio counts on clusters leads to optimistic 
results. This implies that even our results, which respect both NVSS counts and gamma-ray limits, should still be considered rather optimistic.
We note that, owing to our simplified approach using a gamma-ray luminosity--mass relation, these conclusions can be generalised 
to any source of CR protons where these mix and hadronically interact with the ICM of galaxy clusters, such as those injected by structure formation 
shocks and AGNs. In fact, for any considered source of protons, the resulting secondary emission must respect both radio and gamma-ray constraints.

%%%%%%%%%%%%%%%%%%%%%%%%%%%%%%%%%%%%%%%%%%%%%%%%%%%%%%%%
%%%%%%%%%%%%%%%%%%%%%%%%%%%%%%%%%%%%%%%%%%%%%%%%%%%%%%%%
\section{Future detection prospects}
Recently, Ref.~\cite{24} have provided neutrino flux upper limits on a stacked sample of nearby galaxy clusters
(Virgo, Centaurus, Perseus, Coma, and Ophiuchus) following predictions provided by Ref.~\cite{murase2008}. 
These upper limits are $I_{\nu,\,\rmn{UL}}(250~\rmn{TeV}) = 6.9 \times 10^{-20}$ and $7.7 \times 10^{-20}$~cm$^{-2}$~s$^{-1}$~GeV$^{-1}$  
for the cases where CR protons are supposed to be uniformly distributed within the cluster virial radius (A) and where CRs are assumed to be 
distributed like the ICM in clusters (B), respectively. 

The above neutrino flux upper limits were obtained by assuming a spectral index of $\approx 2.15$, so we can compare with our results 
for $\alpha_\rmn{p} = 2.2$. For the stacked sample of Virgo, Centaurus, Perseus, Coma, and Ophiuchus, we estimate that the
upper limits from Ref.~\cite{24} are just a factor of $1.3$ (A) to $1.5$ (B) above the maximum allowed hadronic-induced neutrino 
flux. When $\alpha_\rmn{p} = 2.4$, the maximum allowed flux for the stacked sample is one order of magnitude lower, while for 
$\alpha_\rmn{p} = 2$ it is one order of magnitude higher, with respect to $\alpha_\rmn{p} = 2.2$. 
While we stress that any realistic modelling of these objects should consider their specific characteristic, in particular
the presence or lack of diffuse radio emission, we can conclude that IceCube should be able to put constraints on our 
most optimistic scenario with $\alpha_\rmn{p} = 2$ and on the $\alpha_\rmn{p} = 2.2$ case in the very near future, 
while the case with $\alpha_\rmn{p} = 2.4$ will much harder to test.

%%%%%%%%%%%%%%%%%%%%%%%%%%%%%%%%%%%%%%%%%%%%%%%%%%%%%%%%
%%%%%%%%%%%%%%%%%%%%%%%%%%%%%%%%%%%%%%%%%%%%%%%%%%%%%%%%
\section{Anisotropies in the gamma-ray background}
The authors of Ref.~\cite{25} have shown that a major fraction of the EGB is made by discrete sources
and the measured level of anisotropies is consistent with predictions for gamma-ray blazars. While
we showed that clusters are not the dominant contributors to the the total gamma-ray and neutrino fluxes, 
they could make substantial contributions to the EGB {\it \emph{anisotropies}}. 

\begin{figure}[hbt!]
\centering
\includegraphics[width=0.5\textwidth]{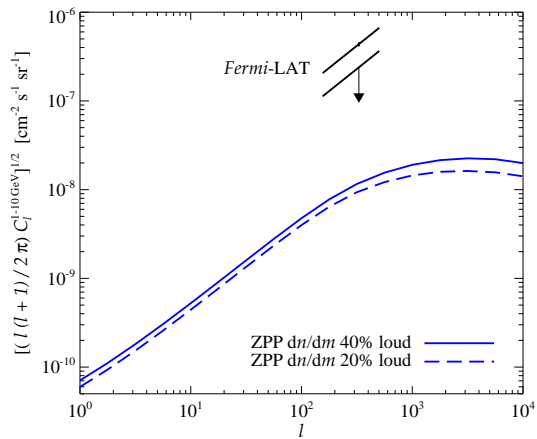}
\caption{Gamma-ray angular power spectrum for emission resulting from proton-proton interactions in galaxy clusters
in the energy range $1-10$~GeV. We show the result for our semi-analytical approach for 20\% and 40\% loud clusters. 
We also show both the measured EGB anisotropy and the upper limits obtained once the blazar component is subtracted \cite{25}.
}
\label{fig:results_ps}
\end{figure}

Figure~\ref{fig:results_ps} shows the angular power spectrum estimated using our semi-analytical 
model and compared with the EGB measurement, and upper limits obtained once the main blazar 
component is subtracted \cite{25}. Our prediction is about one order of magnitude less than the {\it Fermi}-LAT 
upper limit. We underline that in scenarios where the total galaxy cluster intensity is much higher than for our 
semi-analytical model, as is potentially realised for some of our simple phenomenological models based on the 
luminosity--mass relation, the angular power spectrum could be a powerful discriminator, as powerful as radio counts.

%%%%%%%%%%%%%%%%%%%%%%%%%%%%%%%%%%%%%%%%%%%%%%%%%%%%%%%%
%%%%%%%%%%%%%%%%%%%%%%%%%%%%%%%%%%%%%%%%%%%%%%%%%%%%%%%%
\section{Conclusions}
We estimated the contribution from hadronic proton-proton interactions in galaxy clusters to the total extragalactic gamma-ray
and neutrino fluxes, while including radio constraints for the first time. We modelled the cluster population 
by means of their mass function, and make use of a phenomenological luminosity--mass relation applied to all clusters constructed 
by requiring both radio counts and current gamma-ray upper limits to be respected. We also adopted a more refined approach that employs a semi-analytical 
model for CR, ICM and magnetic field distributions in clusters. With these we showed that galaxy clusters can contribute at most up 
to $10\%$ to the neutrino flux measured by IceCube, while contributing much less to the EGB. Nevertheless, IceCube can test the 
most optimistic scenarios for spectral indexes $\geq-2.2$ by stacking few nearby massive objects. We also discussed that galaxy clusters 
could substantially contribute to the EGB anisotropy because they are fewer in number than other astrophysical sources and, therefore, 
are expected to be more anisotropic. Concluding, our results put earlier work, which turned out to be overly optimistic in 
estimating the galaxy cluster contribution, into prospective.

%%%%%%%%%%%%%%%%%%%%%%%%%%%%%%%%%%%%%%%%%%%%%%%%%%%%%%%%
%%%%%%%%%%%%%%%%%%%%%%%%%%%%%%%%%%%%%%%%%%%%%%%%%%%%%%%%

\end{document}